\documentclass{Interspeech2024}
\usepackage{algorithm,algorithmic,cite,colortbl,tabularx}
\hypersetup{colorlinks=true,citecolor=green,linkcolor=red,urlcolor=magenta}
\makeatletter
\newcommand\footnoteref[1]{\protected@xdef\@thefnmark{\ref{#1}}\@footnotemark}
\makeatother
\newcommand{\first}[1]{\cellcolor[rgb]{1,0.7,0.7}{#1}}
\newcommand{\second}[1]{\cellcolor[rgb]{1,0.85,0.7}{#1}}
\newcommand{\third}[1]{\cellcolor[rgb]{1,1,0.7}{#1}}
\interspeechcameraready 

\title{FastVoiceGrad: One-step Diffusion-Based Voice Conversion\\
  with Adversarial Conditional Diffusion Distillation}

\name[affiliation={}]{Takuhiro}{Kaneko}
\name[affiliation={}]{Hirokazu}{Kameoka}
\name[affiliation={}]{Kou}{Tanaka}
\name[affiliation={}]{Yuto}{Kondo}

\address{NTT Corporation, Japan}
\email{takuhiro.kaneko@ntt.com}


\begin{document}

\maketitle

\begin{abstract}
  Diffusion-based voice conversion (VC) techniques such as VoiceGrad have attracted interest because of their high VC performance in terms of speech quality and speaker similarity. However, a notable limitation is the slow inference caused by the multi-step reverse diffusion. Therefore, we propose \textit{FastVoiceGrad}, a novel \textbf{\textit{one-step}} diffusion-based VC that reduces the number of iterations from dozens to \textit{one} while inheriting the high VC performance of the multi-step diffusion-based VC. We obtain the model using \textit{adversarial conditional diffusion distillation (ACDD)}, leveraging the ability of generative adversarial networks and diffusion models while reconsidering the initial states in sampling. Evaluations of one-shot any-to-any VC demonstrate that \textit{FastVoiceGrad} achieves VC performance superior to or comparable to that of previous multi-step diffusion-based VC while enhancing the inference speed.\footnote{\label{foot:samples}Audio samples are available at \url{https://www.kecl.ntt.co.jp/people/kaneko.takuhiro/projects/fastvoicegrad/}.}
\end{abstract}

\section{Introduction}
\label{sec:introduction}

Voice conversion (VC) is a technique for converting one voice into another without changing linguistic contents.
VC began to be studied in a parallel setting, in which mappings between the source and target voices are learned in a supervised manner using a parallel corpus.
However, this approach encounters difficulties in collecting a parallel corpus.
Alternatively, non-parallel VC, which learns mappings without a parallel corpus, has attracted significant interest.
In particular, the emergence of deep generative models has ushered in breakthroughs.
For example, (variational) autoencoder (VAE/AE)~\cite{DKingmaICLR2014}-based VC~\cite{CHsuAPSIPA2016,HKameokaTASLP2019,PTobingIS2019,KTanakaSSW2023,KQianICML2019,JChouIS2019,YHChenICASSP2021,DWangIS2021}, generative adversarial network (GAN)~\cite{IGoodfellowNIPS2014}-based VC~\cite{TKanekoEUSIPCO2018,HKameokaSLT2018,TKanekoIS2019,HKameokaTASLP2020c,TKanekoICASSP2021,YLiIS2021}, flow~\cite{LDinhICLRW2015}-based VC~\cite{JSerraNeurIPS2019}, and diffusion~\cite{JSohlICML2015}-based VC~\cite{HKameokaTASLP2024,SLiuASRU2021,VPopovICLR2022} have demonstrated impressive results.

\begin{figure}[t]  
  \centering
  \includegraphics[width=0.99\linewidth]{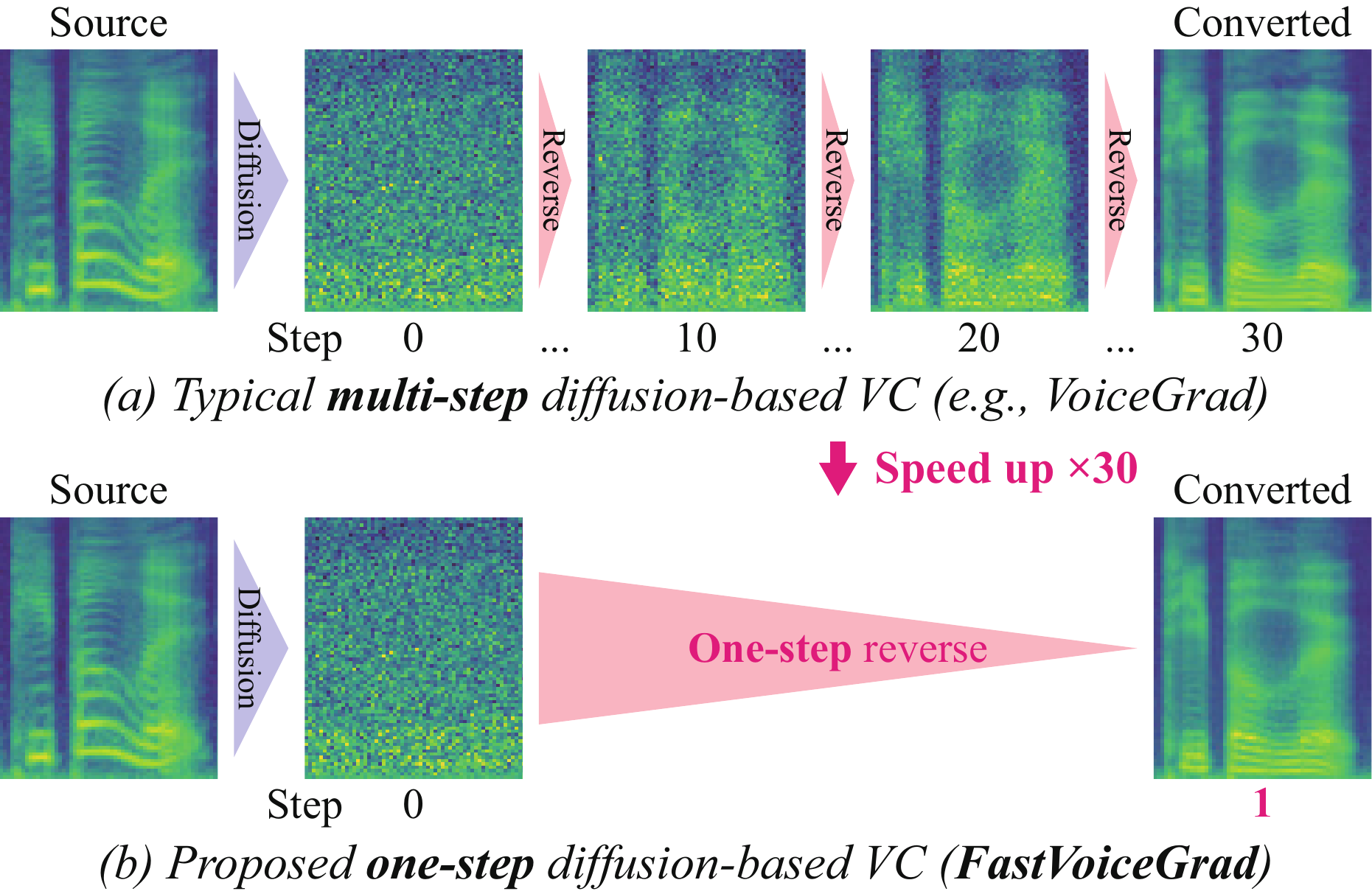}
  \caption{Comparison between (a) typical multi-step diffusion-based VC (e.g., VoiceGrad~\cite{HKameokaTASLP2024}) and (b) proposed \textbf{\textit{one-step}} diffusion-based VC (\textit{FastVoiceGrad}).
    \textit{FastVoiceGrad} reduces the required number of iterations from dozens to \textit{one} and improves the inference speed (e.g., $\times 30$ in this example).}
  \label{fig:teaser}
  \vspace{-2mm}
\end{figure}

Among these models, this paper focuses on diffusion-based VC because it~\cite{HKameokaTASLP2024,VPopovICLR2022} outperforms representative VC models (e.g., \cite{HKameokaTASLP2020c,KQianICML2019,YHChenICASSP2021,DWangIS2021,SLiuTASPL2021}) and has a significant potential for development owing to advancements in diffusion models in various fields (e.g., image synthesis~\cite{YSongNeurIPS2019,JHoNeurIPS2020,ANicholICML2021} and speech synthesis~\cite{NChenICLR2021,ZKongICLR2021}).
Despite these appealing properties, its limitation is the slow inference caused by an iterative reverse diffusion process to transform noise into acoustic features (e.g., the mel spectrogram\footnote{For ease of reading, we hereafter focus on the mel spectrogram as a conversion target but other acoustic features can be equally applied.}) as shown in Figure~\ref{fig:teaser}(a).
This requires at least approximately five iterations, typically dozens of iterations, to obtain sufficiently high-quality speech.
This is disadvantageous compared to other deep generative model-based VC (e.g., VAE-based VC and GAN-based VC discussed above) because they can accomplish VC through a one-step feedforward process.

To overcome this limitation, we propose \textit{FastVoiceGrad}, a novel \textbf{\textit{one-step}} diffusion-based VC model that inherits strong VC capabilities from a multi-step diffusion-based VC model (e.g., VoiceGrad~\cite{HKameokaTASLP2024}), while reducing the required number of iterations from dozens to \textit{one}, as depicted in Figure~\ref{fig:teaser}(b).
To construct this efficient model, we propose \textit{adversarial conditional diffusion distillation (ACDD)}, which is inspired by adversarial diffusion distillation (ADD)~\cite{ASauerArXiv2023} proposed in image synthesis, and distills a multi-step teacher diffusion model into a one-step student diffusion model while exploiting the abilities of GANs~\cite{IGoodfellowNIPS2014} and diffusion models~\cite{JSohlICML2015}.
Note that ADD and ACDD differ in two aspects: (1) ADD addresses a \textit{generation} task (i.e., generating data from \textit{random noise}), while ACDD addresses a \textit{conversion} task (i.e., generating target data from \textit{source data}); and (2) ADD is applied to
\textit{images}, while ACDD is applied to \textit{acoustic features}.
Owing to these differences, we (1) reconsider the initial states in sampling (Section~\ref{subsec:sampling}) and (2) explore the optimal configurations for VC (Section~\ref{subsec:distillation}).

In the experiments, we examined the effectiveness of \textit{FastVoiceGrad} for one-shot any-to-any VC, in which we used an any-to-any extension of VoiceGrad~\cite{HKameokaTASLP2024} as a teacher model and distilled it into \textit{FastVoiceGrad}.
Experimental evaluations indicated that \textit{FastVoiceGrad} outperforms VoiceGrad with the same step (i.e.,  one-step) reverse diffusion process, and has performance comparable to VoiceGrad with a 30-step reverse diffusion process.
Furthermore, we demonstrate that \textit{FastVoiceGrad} is superior to or comparable to DiffVC~\cite{VPopovICLR2022}, another representative diffusion-based VC, while improving the inference speed.

The remainder of this paper is organized as follows:
Section~\ref{sec:voicegrad} reviews VoiceGrad, which is the baseline.
Section~\ref{sec:fastvoicegrad} describes the proposed \textit{FastVoiceGrad}.
Section~\ref{sec:experiments} presents our experimental results.
Finally, Section~\ref{sec:conclusion} concludes the paper with a discussion on future research.

\section{Preliminary: VoiceGrad}
\label{sec:voicegrad}

VoiceGrad~\cite{HKameokaTASLP2024} is a pioneering diffusion-based VC model that includes two variants: a denoising score matching (DSM)~\cite{PVincentNC2011}-based and denoising diffusion probabilistic model (DDPM)~\cite{JHoNeurIPS2020}-based models.
The latter can achieve a VC performance comparable to that of the former while reducing the number of iterations from hundreds to approximately ten~\cite{HKameokaTASLP2024}.
Thus, this study focuses on the DDPM-based model.
The original VoiceGrad was formulated for any-to-many VC.
However, we formulated it for any-to-any VC as a more general formulation.
The main difference is that speaker embeddings are extracted using a speaker encoder instead of speaker labels, while the others remain almost the same.

\smallskip\noindent\textbf{Overview.}
DDPM~\cite{JHoNeurIPS2020} represents a data-to-noise (\textit{diffusion}) process using a gradual nosing process, i.e., $\bm{x}_0 \rightarrow \bm{x}_1 \rightarrow \dots \rightarrow \bm{x}_T$, where $T$ is the number of steps ($T = 1000$ in practice), $\bm{x}_0$ represents real data (mel spectrogram in our case), and $\bm{x}_T$ indicates noise $\bm{x}_T \sim \mathcal{N}(\bm{0}, \bm{I})$.
By contrast, it performs a noise-to-data (\textit{reverse diffusion}) process, that is, $\bm{x}_T \rightarrow \bm{x}_{T-1} \rightarrow \dots \rightarrow \bm{x}_0$, using a gradual denoising process via a neural network.
The details of each process are as follows:

\smallskip\noindent\textit{Diffusion process.}
Assuming a Markov chain, a one-step diffusion process $q(\bm{x}_t | \bm{x}_{t-1})$ ($t \in \{1, \dots, T \}$) is defined as follows:
\begin{flalign}
  \label{eq:one_step_noising}
  q(\bm{x}_t | \bm{x}_{t-1}) = \mathcal{N}(\bm{x}_t; \sqrt{\alpha_t} \bm{x}_{t-1}, \beta_t \mathbf{I}),
\end{flalign}
where $\alpha_t = 1 - \beta_t$.
Owing to the reproductivity of the normal distribution, $q(\bm{x}_t | \bm{x}_0)$ can be obtained analytically as follows:
\begin{flalign}
  \label{eq:total_step_noising}
  q(\bm{x}_t | \bm{x}_0) = \mathcal{N}(\bm{x}_t; \sqrt{\bar{\alpha_t}} \bm{x}_0, (1 - \bar{\alpha_t}) \mathbf{I}),
\end{flalign}
where $\bar{\alpha}_t = \prod_{i=1}^t \alpha_i$.
Using a reparameterization trick~\cite{DKingmaICLR2014}, Equation~(\ref{eq:total_step_noising}) can be rewritten as
\begin{flalign}
  \label{eq:total_step_noising_repara}
  \bm{x}_t = \sqrt{\bar{\alpha_t}} \bm{x}_0 + \sqrt{1 - \bar{\alpha_t}} \bm{\epsilon},
\end{flalign}
where $\bm{\epsilon} \sim \mathcal{N}(\mathbf{0}, \mathbf{I})$.
In practice, $\beta_t$ is fixed at constant values~\cite{JHoNeurIPS2020} with a predetermined noise schedule (e.g., a cosine schedule~\cite{ANicholICML2021}).

\smallskip\noindent\textit{Reverse diffusion process.}
A one-step reverse diffusion process $p_{\theta}(\bm{x}_{t-1} | \bm{x}_t)$ is defined as follows:
\begin{flalign}
  \label{eq:one_step_denoising}
  p_{\theta}(\bm{x}_{t-1} | \bm{x}_t) = \mathcal{N}(\bm{x}_{t - 1}; \bm{\mu}_{\theta}(\bm{x}_t, t, \bm{s}, \bm{p}), \sigma_t^2 \mathbf{I}),
\end{flalign}
where $\bm{\mu}_{\theta}$ indicates the output of a model that is parameterized using $\theta$, conditioned on $t$, speaker embedding $\bm{s}$, and phoneme embedding $\bm{p}$, and $\sigma_t^2 = \frac{1 - \bar{\alpha}_{t - 1}}{1 - \bar{\alpha_t}} \beta_t$.
Unless otherwise specified, $\bm{x}_0$, $\bm{s}$, and $\bm{p}$ are extracted from the same waveform.
Through reparameterization~\cite{DKingmaICLR2014}, Equation~(\ref{eq:one_step_denoising}) can be rewritten as
\begin{flalign}
  \label{eq:one_step_denoising_repara}
  \bm{x}_{t - 1} = \bm{\mu}_{\theta}(\bm{x}_t, t, \bm{s}, \bm{p}) + \sigma_t \bm{z},
\end{flalign}
where $\bm{z} \sim \mathcal{N}(\textbf{0}, \textbf{I})$.

\noindent\textbf{Training process.}
The training objective of DDPM is to minimize the variational bound on the negative log-likelihood $\mathbb{E} [ - \log p_{\theta} (\bm{x}_0) ]$:
\begin{flalign}
  \label{eq:ddpm_loss}
  \mathcal{L}_{\mathrm{DDPM}}(\theta)
  = \mathbb{E}_{q(\bm{x}_{1:T} | \bm{x}_0)} \left[ - \log \frac{p_{\theta}(\bm{x}_{0:T})}{q(\bm{x}_{1:T} | \bm{x}_0)} \right].
\end{flalign}
Using Equation~(\ref{eq:total_step_noising_repara}) and the following reparameterization
\begin{flalign}
  \label{eq:mu_repara}
  \hspace{-2mm}
  \bm{\mu}_{\theta}(\bm{x}_t, t, \bm{s}, \bm{p}) = \frac{1}{\sqrt{\alpha_t}} \left( \bm{x}_t - \frac{1 - \alpha_t}{\sqrt{1 - \bar{\alpha_t}}} \bm{\epsilon}_{\theta}(\bm{x}_t, t, \bm{s}, \bm{p}) \right),
\end{flalign}
Equation~(\ref{eq:ddpm_loss}) can be rewritten as follows:
\begin{flalign}
  \label{eq:ddpm_loss_rewritten}
  \mathcal{L}_{\mathrm{DDPM}}(\theta) = \sum_{t=1}^T w_t \mathbb{E}_{\bm{x}_0, \bm{\epsilon}} [ \lVert \bm{\epsilon} - \bm{\epsilon}_{\theta}(\bm{x}_t, t, \bm{s}, \bm{p}) \rVert_1 ],
\end{flalign}
where $\bm{\epsilon}_{\theta}$ represents a noise predictor that predicts $\bm{\epsilon}$ using $\bm{x}_t$, $t$, $\bm{s}$, and $\bm{p}$.
See~\cite{JHoNeurIPS2020} for detailed derivations of Equations~(\ref{eq:ddpm_loss})--(\ref{eq:ddpm_loss_rewritten}).
Here, $w_t$ is a constant and is set to $1$ in practice for better training~\cite{JHoNeurIPS2020}.
In the original DDPM~\cite{JHoNeurIPS2020}, the L2 loss is used in Equation~(\ref{eq:ddpm_loss_rewritten}); however, we use the L1 loss according to~\cite{NChenICLR2021,HKameokaTASLP2024}, which shows that the L1 loss is better than the L2 loss.

\smallskip\noindent\textbf{Conversion process.}
When $\bm{\epsilon}_{\theta}$ is trained, VoiceGrad can convert the given source mel-spectrogram $\bm{x}_0^{src}$ into a target mel-spectrogram $\bm{x}_0^{tgt}$ using Algorithm~\ref{alg:conversion}.
Here, we use the superscripts ${src}$ and ${tgt}$ to indicate that the data are related to the source and target speakers, respectively.
In this algorithm, a target speaker embedding $\bm{s}^{tgt}$ and a source phoneme embedding $\bm{p}^{src}$ are used as auxiliary information.
To accelerate sampling~\cite{ANicholICML2021}, we use the subsequence $\{ S_K, \dots, S_1 \}$ as a sequence of $t$ values instead of $\{ T, \dots, 1 \}$, where $K \leq T$.
Owing to this change, $\alpha_{S_k}$ is redefined as $\alpha_{S_k} = \frac{\bar{\alpha}_{S_k}}{\bar{\alpha}_{S_{k-1}}}$ for $k > 1$ and $\alpha_{S_k} = \bar{\alpha}_{S_k}$ for $k = 1$.
$\sigma_{S_k}$ is modified accordingly.
Note that VC is a conversion task and not a generation task; therefore, $\bm{x}_0^{src}$ is used as an initial value of $\bm{x}$ (line~1) instead of random noise $\bm{x}_T \sim \mathcal{N}(\textbf{0}, \textbf{I})$, which is typically used in a generation task.
For the same reason, the initial value of $t$ is adjusted from $T$ to $S_K < T$ (line~2) to initiate the reverse diffusion process from the midterm state rather than from the noise.

\begin{algorithm}[t]
  \caption{Conversion process in VoiceGrad~\cite{HKameokaTASLP2024}}
  \label{alg:conversion}
  \begin{algorithmic}[1]
    \renewcommand{\COMMENT}[2][12.5em]{
      \leavevmode\hfill\makebox[#1][l]{$\triangleright$~{\scriptsize #2}}}
    \REQUIRE $\bm{x}_0^{src}$, $\bm{s}^{tgt}$, $\bm{p}^{src}$
    \STATE $\bm{x} \leftarrow \bm{x}_0^{src}$
    \FOR {$t = S_K, \dots, S_1$}
    \STATE $\bm{z} \sim \mathcal{N}(\mathbf{0}, \mathbf{I})$ if $t > S_1$ else $\bm{z} = \textbf{0}$
    \STATE $\bm{x} \leftarrow \frac{1}{\sqrt{\alpha_t}} \left( \bm{x} - \frac{1 - \alpha_t}{\sqrt{1 - \bar{\alpha_t}}} \bm{\epsilon}_{\theta}(\bm{x}, t, \bm{s}^{tgt}, \bm{p}^{src}) \right) + \sigma_t \bm{z}$
    \ENDFOR
    \STATE $\bm{x}_0^{tgt} \leftarrow \bm{x}$
    \RETURN $\bm{x}_0^{tgt}$
  \end{algorithmic}
\end{algorithm}

\section{Proposal: \textit{FastVoiceGrad}}
\label{sec:fastvoicegrad}

\subsection{Rethinking initial states in sampling}
\label{subsec:sampling}

In Algorithm~\ref{alg:conversion}, the two crucial factors that affect the inheritance of source speech are the initial values of (1) $\bm{x}$ and (2) $t$.

\smallskip\noindent\textit{Rethinking the initial value of $\bm{x}$.}
When the initial value of $\bm{x}$ is set to $\bm{x} \sim \mathcal{N}(\mathbf{0}, \mathbf{I})$ (a strategy used in generation), no gap occurs between training and inference; however, we cannot inherit the source information, that is, $\bm{x}_0^{src}$, which is useful for VC to preserve the content.
In contrast, when $\bm{x}_0^{src}$ is directly used as the initial value of $\bm{x}$ (the strategy used in VoiceGrad), we can inherit the source information, but a gap occurs between training and inference.
Considering both aspects, we propose the use of a diffused source mel-spectrogram $\bm{x}_{S_K}^{src}$, defined as
\begin{flalign}
  \label{eq:diffused_input}
  \bm{x}_{S_K}^{src} =  \sqrt{\bar{\alpha}_{S_K}} \bm{x}_0^{src} + \sqrt{1 - \bar{\alpha}_{S_K}} \bm{\epsilon}.
\end{flalign}
In line~1 of Algorithm~\ref{alg:conversion}, $\bm{x}_{S_K}^{src}$ is used instead of $\bm{x}_0^{src}$.
The effect of this replacement is discussed in the next paragraph.

\begin{figure}[t]
  \centering
  \includegraphics[width=0.99\linewidth]{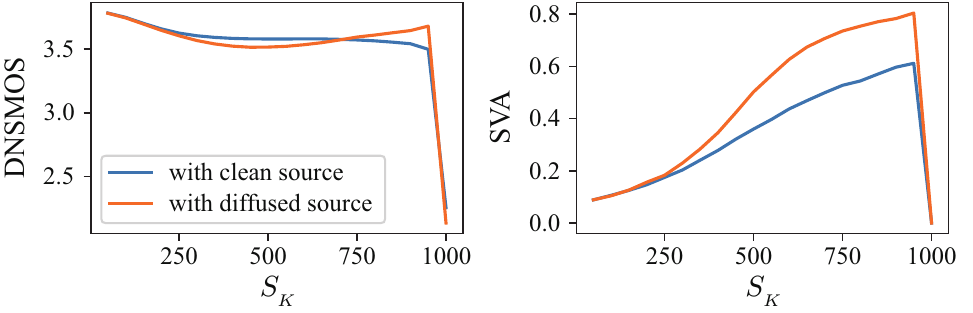}
  \vspace{-1mm}
  \caption{Relationship between DNSMOS and $S_K$ and that between SVA and $S_K$.
    Clean source $\bm{x}_0^{src}$ (blue line) and diffused source $\bm{x}_{S_K}^{src}$ (orange line) were used as initial values of $\bm{x}$.
    The scores were calculated for $S_K$ sampled per 50 steps.}
  \label{fig:initial_state}
  \vspace{-2mm}
\end{figure}

\newpage
\smallskip\noindent\textit{Rethinking the initial value of $t$ (i.e., $S_K$).}
As $S_K$ is closer to $T$, $\bm{x}$ can be transformed to a greater extent under the assumption that it contains more noise, but can corrupt essential information.
As this is a nontrivial tradeoff, it is empirically investigated.
Figure~\ref{fig:initial_state} shows the relationship between $S_K$ and \textit{DNSMOS}~\cite{CReddyICASSP2021} (corresponding to speech quality) and that between $S_K$ and speaker verification accuracy (\textit{SVA})~\cite{BDesplanquesIS2020} (corresponding to speaker similarity).
We present the results for two cases in which $\bm{x}_0^{src}$ and $\bm{x}_{S_K}^{src}$ are used as the initial values of $\bm{x}$.
$K$ was set to $1$; that is, one-step reverse diffusion was conducted.
We observe that SVA improves as $S_K$ increases because $\bm{x}$ is largely transformed toward the target speaker in this case.
When $\bm{x}$ was initialized with $\bm{x}_0^{src}$, DNSMOS worsens as $S_K$ increases.
In contrast, when $\bm{x}$ was initialized with $\bm{x}_{S_K}^{src}$, DNSMOS worsens once but then becomes better, possibly because, in the latter case, the gap between training and inference is alleviated via a diffusion process (Equation~(\ref{eq:diffused_input})) as $S_K$ increases.
Both scores decreasd significantly when $S_K = 1000$, where $\bm{x}$ was denoised under the assumption that $\bm{x}$ is noise.
These results indicate that the one-step reverse diffusion should begin under the assumption that $\bm{x}$ contains the source information, albeit in extremely small amounts.\footnote{Note that, if $K$ is sufficiently large, adequate speech can be obtained even with $S_K = 1000$ at the expense of speed.}
A comparison of the results for $\bm{x}_0^{src}$ and $\bm{x}_{S_K}^{src}$ indicates that $\bm{x}_{S_K}^{src}$ is superior, particularly when considering the SVA.
Based on these results, we used $\bm{x}_{S_K}^{src}$ with $S_K = 950$ in the subsequent experiments.
Figure~\ref{fig:teaser} shows the results for this setting.

\subsection{Adversarial conditional diffusion distillation}
\label{subsec:distillation}

Owing to the difficulty in learning a one-step diffusion model comparable to a multi-step model from scratch, we used a model pretrained using the standard VoiceGrad as an initial model and improved it through ACDD.
Inspired by ADD~\cite{ASauerArXiv2023}, which was proposed for image generation, we used adversarial loss and score distillation loss in distillation.

\smallskip\noindent\textbf{Adversarial loss.}
Initially, we considered directly applying a discriminator to the mel spectrogram, similar to the previous GAN-based VC (e.g.,~\cite{TKanekoICASSP2021,YLiIS2021}).
However, we could not determine an optimal discriminator to eliminate the buzzy sound in the waveform.
Therefore, we converted the mel spectrogram into a waveform using a neural vocoder $\mathcal{V}$ (with frozen parameters) and applied a discriminator $\mathcal{D}$ in the waveform domain.
More specifically, adversarial loss (particularly least-squares GAN~\cite{XMaoICCV2017}-based loss) is expressed as follows:
\begin{flalign}
  \label{eq:adv_loss}
  \mathcal{L}_{\mathrm{adv}}(\mathcal{D})
  & = \mathbb{E}_{\bm{x}_0} [ (\mathcal{D}(\mathcal{V}(\bm{x}_0)) - 1)^2
    + (\mathcal{D}(\mathcal{V}(\bm{x}_{\theta})))^2],
  \\
  \mathcal{L}_{\mathrm{adv}}(\theta)
  & = \mathbb{E}_{\bm{x}_0} [ (\mathcal{D}(\mathcal{V}(\bm{x}_{\theta})) - 1)^2 ],
\end{flalign}
where $\bm{x}_0$ represents a mel spectrogram extracted from real speech.
$\bm{x}_{\theta}$ represents a mel spectrogram generated using $\bm{x}_{\theta} = \bm{\mu}_{\theta}(\bm{x}_{S_k}, S_K, \bm{s}, \bm{p})$ (one-step denoising prediction defined in Equation~(\ref{eq:mu_repara})), where $\bm{x}_{S_K}$ is the $S_K$-step diffused $\bm{x}_0$ via Equation~(\ref{eq:diffused_input}).
The adversarial loss is used to improve the reality of $\bm{x}_{\theta}$ through adversarial training.

Furthermore, following the training of a neural vocoder~\cite{KKumarNeurIPS2019,JKongNeurIPS2020}, we used the feature matching (FM) loss, defined as
\begin{flalign}
  \label{eq:fm_loss}
  \mathcal{L}_{\mathrm{FM}}(\theta)
  = \mathbb{E}_{\bm{x}_0}\left[ \sum_{l = 1}^L \frac{1}{N_l} \lVert \mathcal{D}_l(\mathcal{V}(\bm{x}_0)) - \mathcal{D}_l(\mathcal{V}(\bm{x}_{\theta})) \rVert_1 \right],
\end{flalign}
where $L$ indicates the number of layers in $\mathcal{D}$.
$\mathcal{D}_l$ and $N_l$ denote the features and the number of features in the $l$-th layer of $\mathcal{D}$, respectively.
$\mathcal{L}_{\mathrm{FM}}(\theta)$ bears $\bm{x}_{\theta}$ closer to $\bm{x}_0$ in the discriminator feature space.

\smallskip\noindent\textbf{Score distillation loss.}
The score distillation loss~\cite{ASauerArXiv2023} is formulated as follows:
\begin{flalign}
  \label{eq:distill_loss}
  \mathcal{L}_{\mathrm{dist}}(\theta)
  = \mathbb{E}_{t, \bm{x}_0} [ c(t) \lVert \bm{x}_{\phi} - \bm{x}_{\theta} \rVert_1 ],
\end{flalign}
where $\bm{x}_{\phi}$ is one-step denoising prediction (Equation~(\ref{eq:mu_repara})) generated by a teacher diffusion model parameterized with $\phi$ (frozen in training): $\bm{x}_{\phi} = \bm{\mu}_{\phi}(\mathrm{sg}(\bm{x}_{\theta, t}), t, \bm{s}, \bm{p})$.
Here, $\mathrm{sg}$ denotes the stop-gradient operation, $\bm{x}_{\theta, t}$ is the $t$-step diffused $\bm{x}_{\theta}$ via Equation~(\ref{eq:total_step_noising_repara}), and $t \in \{ 1, \dots, T \}$.
$c(t)$ is a weighting term and is set to $\alpha_t$ in practice to allow higher noise levels to contribute less~\cite{ASauerArXiv2023}.
$\mathcal{L}_{\mathrm{dist}}(\theta)$ encourages $\bm{x}_{\theta}$ (student output) to match $\bm{x}_{\phi}$ (teacher output).

\smallskip\noindent\textbf{Total loss.}
The total loss is expressed as follows:
\begin{flalign}
  \hspace{-2mm}
  \mathcal{L}_{\mathrm{ACDD}}(\theta) & = \mathcal{L}_{\mathrm{adv}}(\theta) + \lambda_{\mathrm{FM}} \mathcal{L}_{\mathrm{FM}}(\theta) + \lambda_{\mathrm{dist}} \mathcal{L}_{\mathrm{dist}} (\theta),
  \\
  \hspace{-2mm}
  \mathcal{L}_{\mathrm{ACDD}}(\mathcal{D}) & = \mathcal{L}_{\mathrm{adv}}(\mathcal{D}),
\end{flalign}
where $\lambda_{\mathrm{FM}}$ and $\lambda_{\mathrm{dist}}$ are weighting hyperparameters set to $2$ and $45$, respectively, in the experiments.
${\theta}$ and $\mathcal{D}$ are optimized by minimizing $\mathcal{L}_{\mathrm{ACDD}}(\theta)$ and $\mathcal{L}_{\mathrm{ACDD}}(\mathcal{D})$, respectively.

\section{Experiments}
\label{sec:experiments}

\subsection{Experimental settings}
\label{subsec:experimental_settings}

\textbf{Data.}
We examined the effectiveness of \textit{FastVoiceGrad} on one-shot any-to-any VC using the VCTK dataset~\cite{JYamagishiVCTK2019}, which included the speeches of 110 English speakers.
To evaluate the unseen-to-unseen scenarios, we used 10 speakers and 10 sentences for testing, whereas the remaining 100 speakers and approximately 390 sentences were used for training.
Following DiffVC~\cite{VPopovICLR2022}, audio clips were downsampled at 22.05kHz, and 80-dimensional log-mel spectrograms were extracted from the audio clips with an FFT size of 1024, hop length of 256, and window length of 1024.
These mel spectrograms were used as conversion targets.

\smallskip\noindent\textbf{Comparison models.}
We used VoiceGrad~\cite{HKameokaTASLP2024} (Section~\ref{sec:voicegrad}) as the main baseline and distilled it into \textit{FastVoiceGrad}.
A diffusion model has a tradeoff between speed and quality according to the number of reverse diffusion steps ($K$).
To investigate this effect, we examined three variants: \textit{VoiceGrad-1}, \textit{VoiceGrad-6}, and \textit{VoiceGrad-30}, which are VoiceGrad with $K = 1$, $K = 6$, and $K = 30$, respectively.
\textit{VoiceGrad-1} is as fast as \textit{FastVoiceGrad}, whereas the others are slower.
For an ablation study, we examined \textit{FastVoiceGrad$_{\mathrm{adv}}$} and \textit{FastVoiceGrad$_{\mathrm{dist}}$}, in which score distillation and adversarial losses were ablated, respectively.
As another strong baseline, we examined DiffVC~\cite{VPopovICLR2022}, which has demonstrated superior quality compared to representative one-shot VC models~\cite{YHChenICASSP2021,DWangIS2021,SLiuTASPL2021}.
Based on~\cite{VPopovICLR2022}, we used two variants: \textit{DiffVC-6} and \textit{DiffVC-30}, that is, DiffVC with six and 30 reverse diffusion steps, respectively.

\smallskip\noindent\textbf{Implementation.}
VoiceGrad and \textit{FastVoiceGrad} were implemented while referring to~\cite{HKameokaTASLP2024}.
We implemented $\bm{\epsilon}_{\theta}$ using U-Net~\cite{ORonnebergerMICCAI2015}, which consisted of 12 one-dimensional convolution layers of 512 hidden channels with two downsampling/upsampling, gated linear unit (GLU) activation~\cite{YDauphinICML2017}, and weight normalization~\cite{TSalimansNIPS2016}.
The two main changes from~\cite{HKameokaTASLP2024} were that speaker embedding $\bm{s}$ was extracted by a speaker encoder~\cite{YJiaNeurIPS2018} instead of a speaker label, and $t$ was encoded by sinusoidal positional embedding~\cite{AVaswaniNIPS2017} instead of one-hot embedding.
We extracted $\bm{p}$ using a bottleneck feature extractor (BNE)~\cite{SLiuTASPL2021}.
We implemented $\mathcal{V}$ and $\mathcal{D}$ using the modified HiFi-GAN-V1~\cite{JKongNeurIPS2020}, in which a multiscale discriminator~\cite{KKumarNeurIPS2019} was replaced with a multiresolution discriminator~\cite{WJangIS2021} that showed better performance in speech synthesis~\cite{WJangIS2021}.
We trained VoiceGrad using the Adam optimizer~\cite{DPKingmaICLR2015} with a batch size of 32, learning rate of 0.0002, and momentum terms $(\beta_1, \beta_2) = (0.9, 0.999)$ for 500 epochs.
We trained \textit{FastVoiceGrad} using the Adam optimizer~\cite{DPKingmaICLR2015} with a batch size of 32, learning rate of 0.0002, and momentum terms $(\beta_1, \beta_2) = (0.5, 0.9)$ for 100 epochs.
We implemented DiffVC using the official code.\footnote{\url{https://github.com/huawei-noah/Speech-Backbones/tree/main/DiffVC}}

\smallskip\noindent\textbf{Evaluation.}
We conducted mean opinion score (MOS) tests to evaluate perceptual quality.
We used 90 different speaker/sentence pairs for the subjective evaluation.
For the speech quality test (\textit{qMOS}), nine listeners assessed the speech quality on a five-point scale: 1 = bad, 2 = poor, 3 = fair, 4 = good, and 5 = excellent.
For the speaker similarity test (\textit{sMOS}), ten listeners evaluated speaker similarity on a four-point scale: 1 = different (sure), 2 = different (not sure), 3 = same (not sure), and 4 = same (sure), in which the evaluated speech was played after the target speech (with a different sentence).
As objective metrics, we used \textit{UTMOS}~\cite{TSaekiIS2022}, \textit{DNSMOS}~\cite{CReddyICASSP2021}, and character error rate (\textit{CER})~\cite{ABaevskiNeurIPS2020} to evaluate speech quality.
We used DNSMOS (MOS sensitive to noise) in addition to UTMOS (which achieved the highest score in the VoiceMOS Challenge 2022~\cite{WHHuangIS2022}) because we found that UTMOS is insensitive to speech with noise, which typically occurs when using a diffusion model with a few reverse diffusion steps.
We evaluated speaker similarity using \textit{SVA}~\cite{BDesplanquesIS2020}, in which we verified whether converted and target speech are uttered by the same speaker.
We used 8,100 different speaker/sentence pairs for objective evaluation.
The audio samples are available from the link indicated on the first page of this manuscript.\footnoteref{foot:samples}

\subsection{Experimental results}
\label{subsec:experimental_results}

Table~\ref{tab:results_vctk} summarizes these results.
We observed that \textit{FastVoiceGrad} not only outperformed the ablated FastVoiceGrads (\textit{FastVoiceGrad$_{\mathrm{adv}}$} and \textit{FastVoiceGrad$_{\mathrm{dist}}$}) and \textit{VoiceGrad-1}, which have the same speed, but was also superior to or comparable to \textit{VoiceGrad-6} and \textit{VoiceGrad-30}, of which calculation costs were as six and 30 times as \textit{FastVoiceGrad}, respectively.
Furthermore, \textit{FastVoiceGrad} was superior to or comparable to DiffVCs (\textit{DiffVC-6} and \textit{DiffVC-30}) in terms of all metrics.\footnote{On the Mann--Whitney U test ($p$-value $> 0.05$), \textit{FastVoiceGrad} is \textit{not} significantly different from \textit{VoiceGrad-30/6} and \textit{DiffVC-30} but significantly better than the other baselines for qMOS, and \textit{FastVoiceGrad} is significantly better than all baselines for sMOS.}
For a single A100 GPU, the real-time factors of mel-spectrogram conversion and total VC (including feature extraction and waveform synthesis) for \textit{FastVoiceGrad} are 0.003 and 0.060, respectively, which are faster than those for \textit{DiffVC-6} (fast variant), which are 0.094 and 0.135, respectively.
These results indicate that \textit{FastVoiceGrad} can enhance the inference speed while achieving high VC performance.

\subsection{Application to another dataset}
\label{subsec:application}

\begin{table}[t]
  \caption{Comparison of qMOS with 95\% confidence interval, sMOS with 95\% confidence interval, UTMOS, DNSMOS, CER [\%], and SVA [\%] for VCTK.}
  \vspace{-2mm}
  \label{tab:results_vctk}
  \newcommand{\spm}[1]{{\tiny$\pm$#1}}
  \setlength{\tabcolsep}{1.2pt}
  \centering
  \scriptsize{
    \begin{tabularx}{\columnwidth}{lcccccc}
      \toprule
      \multicolumn{1}{c}{\textbf{Model}} & \textbf{qMOS$\uparrow$} & \textbf{sMOS$\uparrow$} & \textbf{UTMOS$\uparrow$} & \textbf{DNSMOS$\uparrow$} & \textbf{CER$\downarrow$} & \textbf{SVA$\uparrow$}
      \\ \midrule
      Ground truth
      & 4.24\spm{0.11} & 3.47\spm{0.12} & 4.14 & 3.75 & 1.21 & 100.0
      \\ \midrule
      DiffVC-6
      & 3.34\spm{0.12} & 2.29\spm{0.14} & 3.80 & 3.68 & 6.23 & 65.0
      \\
      DiffVC-30
      & 3.69\spm{0.11} & 2.28\spm{0.14} & 3.76 & 3.75 & 6.84 & 66.1
      \\ \midrule
      VoiceGrad-1
      & 3.00\spm{0.10} & 2.27\spm{0.15} & 3.72 & 3.68 & \third{2.11} & 80.4
      \\
      VoiceGrad-6
      & \third{3.74\spm{0.10}} & 2.26\spm{0.16} & \third{3.93} & 3.74 & 2.13 & \third{81.5}
      \\
      VoiceGrad-30
      & \first{3.95\spm{0.11}} & \second{2.42\spm{0.16}} & 3.88 & \second{3.77} & 2.20 & \second{82.9}
      \\ \midrule
      FastVoiceGrad
      & \second{3.86\spm{0.09}} & \first{2.68\spm{0.16}} & \second{3.96} & \second{3.77} & \first{1.89} & \first{83.0}
      \\
      FastVoiceGrad$_{\mathrm{adv}}$
      & 3.47\spm{0.12} & \third{2.30\spm{0.15}} & 3.62 & \first{3.81} & 2.96 & 72.7
      \\
      FastVoiceGrad$_{\mathrm{dist}}$
      & 3.07\spm{0.11} & 2.11\spm{0.14} & \first{3.98} & 3.67 & \second{2.01} & 76.7
      \\ \bottomrule
    \end{tabularx}
  }
  \vspace{-0.5mm}
\end{table}

\begin{table}[t]
  \caption{Comparison of UTMOS, DNSMOS, CER [\%], and SVA [\%] for LibriTTS.
    $^\dag$Ground-truth converted speech does not necessarily exist in LibriTTS; therefore, alternatively, source speech was used for evaluation.}
  \vspace{-2mm}
  \label{tab:results_libritts}
  \newcommand{\spm}[1]{{\tiny$\pm$#1}}
  \setlength{\tabcolsep}{7pt}
  \centering
  \scriptsize{
    \begin{tabularx}{\columnwidth}{lcccc}
      \toprule
      \multicolumn{1}{c}{\textbf{Model}} & \textbf{UTMOS$\uparrow$} & \textbf{DNSMOS$\uparrow$} & \textbf{CER$\downarrow$} & \textbf{SVA$\uparrow$}
      \\ \midrule
      Ground truth$^\dag$
      & 4.06 & 3.70 & 0.87 & --
      \\ \midrule
      DiffVC-6
      & 3.57 & 3.54 & 2.26 & 77.5
      \\
      DiffVC-30
      & 3.65 & 3.68 & 2.53 & 77.2
      \\ \midrule
      VoiceGrad-1
      & 3.07 & 3.29 & \third{1.37} & 76.2
      \\
      VoiceGrad-6
      & \third{3.83} & 3.67 & 1.44 & \second{78.6}
      \\
      VoiceGrad-30
      & 3.77 & \second{3.74} & 1.52 & 77.8
      \\ \midrule
      FastVoiceGrad
      & \second{3.94} & \first{3.75} & \first{1.31} & \first{80.0}
      \\
      FastVoiceGrad$_{\mathrm{adv}}$
      & 3.48 & \second{3.74} & 1.74 & 73.9
      \\
      FastVoiceGrad$_{\mathrm{dist}}$
      & \first{4.03} & 3.53 & \second{1.33} & \third{78.1}
      \\ \bottomrule
    \end{tabularx}
  }
  \vspace{-2mm}
\end{table}

To confirm this generality, we evaluated \textit{FastVoiceGrad} on the LibriTTS dataset~\cite{HZenIS2019}.
We used the same networks and training settings as those for the VCTK dataset, except that the training epochs for \textit{VoiceGrad} and \textit{FastVoiceGrad} were reduced to 300 and 50, respectively, owing to an increase in the amount of training data.
Table~\ref{tab:results_libritts} summarizes the results.
The same tendencies were observed in that \textit{FastVoiceGrad} not only outperformed \textit{VoiceGrad-1} (a model with the same speed) but was also superior to or comparable to the other baselines.

\section{Conclusion}
\label{sec:conclusion}

We proposed \textit{FastVoiceGrad}, a \textbf{\textit{one-step}} diffusion-based VC model that can achieve VC performance comparable to or superior to multi-step diffusion-based VC models while reducing the number of iterations to \textit{one}.
The experimental results demonstrated the importance of carefully setting of the initial states in sampling and the necessity of the joint use of GANs and diffusion models in distillation.
Future research should include applications to advanced VC tasks (e.g., emotional VC and accent correction) and an extension to real-time implementation.

\section{Acknowledgements}
This work was supported by JST CREST Grant Number JPMJCR19A3, Japan.

\clearpage
\bibliographystyle{IEEEtran}
\bibliography{refs}

\end{document}